\documentclass[aps,english,twocolumn]{revtex4}
\usepackage{mathrsfs}
\usepackage{graphicx}
\usepackage{amsmath, amssymb}
\usepackage{babel}

\begin{document}

\title{ Non-Fermi liquid behavior in Bose-Fermi mixtures at two dimensions}
\author{Xiao-Yong  Feng$^{1}$  and Tai-Kai Ng$^{2}$ }
\affiliation{Condensed Matter Group,  Department of Physics,
Hangzhou Normal University, Hangzhou 310036, China$^1$\\
Department of Physics, Hong Kong University of Science and
Technology, Clear Water Bay Road, Kowloon, Hong Kong, China$^2$}

\begin{abstract}
  In this paper we study the low temperature behaviors of a system of Bose-Fermi mixtures at two dimensions. Within a
  self-consistent ladder diagram approximation, we show that at nonzero temperatures $T\rightarrow0$ the fermions
  exhibit non-fermi liquid behavior. We propose that this is a general feature of Bose-Fermi mixtures at two dimensions.
  An experimental signature of this new state is proposed.
\end{abstract}

\maketitle

  The issue of Bose-Fermi mixture(BFm) can be traced back
to the study of mixture of Helium-4 and Helium-3 \cite{Cohen}.
$^{4}$He atoms carry integer spin
 and obey Bose-Einstein statistics whereas $^{3}$He atoms carry half-integer spin and obey Fermi-Dirac statistics. A natural question is what physics a \
 system of mixture of these two types of particles will display. It was found that the critical temperature of the $^{4}$He condensate is suppressed with
 increasing the concentration of $^{3}$He before phase separation\cite{Abraham,Walters}. Similar phenomena also occur in atomic
 BFm\cite{Modugno,Kenneth,Hebert} with richer physics. Through exchanging Bogoliubov excitons in Bose condensate, fermions gain
 effective attraction between themselves\cite{Fabrizio}, leading to a rich possibility of different orders\cite{Klironomos}. In strong coupling limit,
 composite fermions comprising of one fermion and several bosons or bosonic holes may form, the residual interactions between these
 composite fermions, according to the system parameters, will result in different states such as a density wave, a superfluid
 liquid and an insulator with fermionic domains\cite{Lewenstein}. The Bose-Fermi mixture is predicted to collapse when boson-fermion interaction is
 attractive and strong\cite{Molmer, zhang}.

 Most existing theoretical works consider three-dimensional systems. In this paper we study Bose-Fermi mixtures in two-dimensions.
% The rapid development of experimental technique in cold atom experiments enables us to tune the strength of interaction between trapped
% atoms\cite{Greiner, Zwierlein, Truscott, K}.
  Within a self-consistent ladder-diagram approximation we show that the fermions may form a non-Fermi liquid state because of the unique
  feature that Bosons cannot bose-condense in two dimensions. The non-Fermi liquid state is a result of non-negligible thermal fluctuations of (2D) bosons
  which dominates the boson behavior, leading to non-Fermi liquid behavior of fermions even at $T\rightarrow0$
  and is independent of the sign of fermion-boson interaction.

 We consider a system of bosons and (spinless) fermions described by the Hamiltonian,
 \begin{eqnarray}
 \label{H}
 H & = & \sum_{\vec{q}}\varepsilon_{\vec{q}}b^{\dag}_{\vec{q}}b_{\vec{q}}
 +\sum_{\vec{k}}\xi_{\vec{k}}f^{\dag}_{\vec{k}}f_{\vec{k}}
 \\ \nonumber
 & & -U_0\int d^dx n^{b}(\vec{x})n^{f}(\vec{x})+U_b\int d^dx(n^{b}(\vec{x}))^2
\end{eqnarray}
 where $b$ and $f$ represent boson and fermion, respectively, $\varepsilon_{\vec{q}}=\vec{q}^2/2m_{b}-\mu_b$ and $\xi_{\vec{k}}=\vec{k}^2/2m_{f}-\mu_f$.
 $\mu_{b(f)}, m_{b(f)}$ are the chemical potentials and masses for bosons (fermions), respectively and $n_{b(f)}(\vec{x})$ are the corresponding density
 operators. $U_b$ is the direct interaction between bosons and $U_0$ is the interaction strength between bosons and fermions. We set $\hbar=1$ in this paper.
 We shall consider weak interactions $U_b$ and $U_0$ in the following.
 %We note that the real interaction seen by bosons is a sum of the direct interaction $U_b$ and a fermion-mediated interaction
 %$U_I\sim -N_FU_0^2$, where $N_f\sim n_f/E_f$ is the density of states of fermions on the Fermi surface, $n_f$ and $E_f$ are the
 %fermion density and fermi energy of the fermion system, respectively.

  To treat boson-fermion interaction we go to momentum space and introduce a fermion (Grassman) Hubbard-Strotonovich field $c_{k}$ which describes
  fermion-boson pairing\cite{maeda}, i.e.
 \begin{eqnarray}
  & & \int d^dx n^{b}(\vec{x})n^{f}(\vec{x})=\sum_{k,q,q'}f^{\dag}_{k-q}f_{k-q'}b^{\dag}_{q}b_{q'}
  \nonumber\\
 & \rightarrow &
  \sum_{k,q}f^{\dag}_{k-q}b^+_{q}c_{k}+\sum_{k,q}c^{\dag}_{k}b_{q}f_{k-q}
  -\sum_{k} c^+_{k}c_{k},
  \end{eqnarray}
  leading to an effective BFm action
  \begin{subequations}
\begin{eqnarray}
S &=& S_b+\sum_{k}(-i\omega+\xi_{\vec{k}})f^{+}_{k}f_{k}\nonumber\\
&&+U_0\sum_{k}c^{+}_{k}c_{k}-\frac{U_0}{\sqrt{N\beta}}\sum_{k,q}(f^{+}_{k-q}b^+_{q}c_{k}+c.c.)
\end{eqnarray}
 where
 \begin{equation}
 \label{baction}
 S_b=\sum_{q}(-i\Omega+\varepsilon_{\vec{q}})b^{+}_{q}b_{q}+U_b\int d^dx(n_b(\vec{x}))^2
 \end{equation}
 \end{subequations}
 is the (pure) boson action. $q=(\vec{q},i\Omega)$ and $k=(\vec{k},i\omega)$ with $\Omega=2n\pi/\beta$ and $\omega=(2n+1)\pi/\beta$ in which
 $n=$ integers and $\beta=(k_BT)^{-1}$. $c_{k}$'s describe pairing of fermions and bosons and are the analogue of pairing order parameters
 $\Delta_k$ in BCS theory for superconductors. The major difference between the two situations is that $c_k$'s are Grassman numbers here. As a result, we cannot have
 $\langle c_k\rangle\neq0$ in boson-fermion mixture and a more elaborated technique beyond BCS mean-field theory is needed to treat the $c_k$
 fields.
 %We shall treat $S$ as an effective action describing the low energy physics of the boson-fermion mixture in the following. In this sense
 %$\xi^f_{\vec{k}}$ should be considered as an effective quasi-particle dispersion where the high-energy interaction effects (for example, an
 %Hartree self-energy $U_ox$, $x=$ boson density) are already absorbed, and $U_b$ should be considered as an effective interaction between bosons which
 %includes also fermion-mediated interactions.

\subsubsection{non-interacting bosons}
 To proceed further we first consider non-interacting fermions, i.e. $U_b=0$. In this case we can integrate out the bosons in $S$ straightforwardly since it
 is quadratic in boson fields. We obtain
\begin{eqnarray}
\label{seff}
 S_{eff}&=&\sum_{k}(-i\omega+\xi_{\vec{k}})f^{+}_{k}f_{k}+U_0\sum_{k}c^{+}_{k}c_{k}\nonumber\\
 &&-\frac{U_0^{2}}{N\beta}\sum_{q,k,k'}f^{+}_{k-q}f_{k'-q}c^{+}_{k'}c_{k}\times g_b(q)
\end{eqnarray}
 where $g_b(q)=1/(i\Omega-\varepsilon^b_{\vec{q}})$ is the non-interacting boson green's function.

 To proceed we perform a mean-field decoupling to the fermion-fermion interaction term in
   $S_{eff}$,
  \[
    f^{*}fc^{*}c\sim \langle f^{*}f\rangle
    c^{*}c+f^{*}f\langle c^{*}c\rangle-\langle f^{*}f\rangle\langle
    c^{*}c\rangle,
  \]
   where $\langle f^{*}f\rangle(\langle c^{*}c\rangle)=G_f(k)(G_c(k))$ are $f-$ and $c-$ fermion Green's functions but not
   c-numbers. The mean-field theory leads to the self-consistent equations
   \begin{eqnarray}
  \label{sce}
  G_c(k) & = & \frac{-1}{U_0+U_0^{2}\Pi_c(k)}  \\ \nonumber
  G_f(k) & = & \frac{1}{i\omega-\xi^f_{\vec{k}}-U_0^2\Pi_f(k)}
  \end{eqnarray}
  where
  \begin{eqnarray}
  \label{pis}
  \Pi_c(k) & = & {-1\over\beta N}\sum_qG_f(k-q)g_b(q),  \\ \nonumber
  \Pi_f(k) & = & {-1\over\beta N}\sum_qG_c(k+q)g_b(q).
  \end{eqnarray}
   $G_c$ and $G_f$ are determined self-consistently from Eq.\ (\ref{sce}) and\ (\ref{pis}) and can be viewed as the generalization of BCS theory to
   the case of fermion-boson binding.

   To gain insight to the equations we first examine the fermion-boson bound states in the lowest order ladder diagram approximation where we approximate
  $G_f(k-q)\sim g_{f}(k-q)$ in Eq.\ (\ref{sce}), where $g_f(k)$ is the free fermion green's function\cite{maeda}. In this approximation,
  $\Pi_c(k)\rightarrow\Pi_{c}^{(0)}(k)$, where
  \[
  \Pi_c^{(0)}(k)={-1\over\beta N}\sum_qg_f(k-q)g_b(q)  \]
   is the (bare) fermion-boson bubble diagram. The feedback effects of fermion-boson bound state on the
 fermion green's function $g_f$ is missing in this approximation.

   Performing the sum over frequencies, we obtain $\Pi_c^{(0)}(k, i\omega)=\Pi_{0B}(k, i\omega)+\Pi_{0F}(k, i\omega)$, where
   \begin{subequations}
   \label{pi00}
  \begin{eqnarray}
  \label{pi0}
  \Pi_{0B}(k, i\omega) & = & \int {d^dq\over(2\pi)^d}{n_B(\varepsilon_{\vec{q}})\over
  i\omega-\varepsilon_{\vec{q}}-\xi_{\vec{k}-\vec{q}}}
   \\ \nonumber
  \Pi_{0F}(k, i\omega) & = & \int
  {d^dq\over(2\pi)^d}{1-n_F(\xi_{\vec{k}-\vec{q}})\over
  i\omega-\varepsilon_{\vec{q}}-\xi_{\vec{k}-\vec{q}}}.
  \end{eqnarray}
   It is straightforward to show that in two dimensions,
   \begin{equation}
   \Pi_{0F}(k\rightarrow k_F,\omega\rightarrow\xi_{\vec{k}})\rightarrow
   C-{im_f\over4\pi\varepsilon_{\vec{k}}}\sqrt{4\varepsilon_{\vec{k}}\omega-(\omega-\xi_{\vec{k}})^2}
 \end{equation}
   in the low temperature limit $\beta^{-1}\rightarrow0$, where $C\sim m_f$ is a constant, i.e. $\Pi_F(k\rightarrow k_F,\omega\rightarrow0)$ is a
   non-singular function at the Fermi surface. The situation is quite different for $\Pi_B$. Approximating
   $n_B(\varepsilon)\sim 0$ for $\varepsilon>>\beta^{-1}$ and $n_B(\varepsilon)\sim 1/\beta\varepsilon$ for
   $\varepsilon<<\beta^{-1}$, we obtain
  \begin{equation}
  \label{piB0}
  \Pi_{0B}(k, i\omega) \sim {1\over4\pi^2\beta}\int^{\beta\varepsilon_q<1}
  d^2q{1\over\varepsilon_q(i\omega-\varepsilon_{\vec{q}}-\xi_{\vec{k}-\vec{q}})},
  \end{equation}
  the integral is diverging logarithmically at small $q$ at any finite temperature $k_BT>>\Delta\varepsilon$, where $\Delta\varepsilon$ is the (boson)
  energy level spacing. This singular behavior is a particular feature of (free) boson propagators in two dimensions and is missing in three-dimension
  systems\cite{maeda}. The leading effect of the $q\rightarrow0$ singularity can be extracted by expanding the function
  $(i\omega-\varepsilon_{\vec{q}}-\xi_{\vec{k}-\vec{q}})^{-1}$ in a power series in $q$. Performing the expansion, we obtain
  \begin{equation}
  \label{piB1}
  \Pi_{0B}(k, i\omega) \sim g_f(k)\left(x+O\left[m_bk_BT\ln\left({(i\omega-\xi_{\vec{k}})^2m_f\beta\over
  \varepsilon_{\vec{k}}m_b}\right)\right]\right)
  \end{equation}
  \end{subequations}
  %for $\omega-\xi_{\vec{k}}\geq\beta^{-1}$
  where $x$ is the density of bosons.  We have used the result
  \[ x=\int {d^2q\over(2\pi)^2}n_B(\varepsilon^b_q)\sim {1\over4\pi^2\beta}\int^{\beta\varepsilon^b_q<1} {d^2q\over\varepsilon^b_q},   \]
  in deriving\ (\ref{piB1}), which is valid for non-interacting bosons at two dimensions where Bose-Einstein condensation is absent at any finite temperature.
   The singular behavior of the integral permits us to extract the leading contribution
   to $\Pi_{0B}(k,i\omega)$ analytically at $T\rightarrow0$.

   We now return to Eq.\ (\ref{sce}). The self-consistent equations are difficult to solve in general. However, the leading singular contribution at $q\rightarrow0$ at two dimensions to the
   propagators $\Pi_c$ and $\Pi_f$ at $k\rightarrow k_F$, $\omega\rightarrow0$ can be extracted in the
   limit $\beta^{-1}\rightarrow0$ rather straightforwardly.
  % The idea is that as far as extracting the leading singular contributions at $\vec{q}\rightarrow0$ is concerned, we may
  % approximate
  % \begin{eqnarray}
  % \sum_qG(k-q)g_b(q) & = & \sum_qg_b(q)\left(I-q.\partial_k+...\right)G(k)
  % \\ \nonumber
  % & \sim & xG(k)
  % \end{eqnarray}
  %where $G(k)=G_{c(f)}(k)$. One may check the validity of the above approximation directly by
  Using the spectral representation
   \[
       G_{c(f)}(\vec{k},i\omega)={1\over\pi}\int d\epsilon{A_{c(f)}(\vec{k},\epsilon)\over i\omega-\epsilon},
       \]
   we can perform the sum over frequencies in $\Pi_c$ and $\Pi_f$ and $\Pi_{c(f)}$ can be separated into
  $\Pi_{c(f)}=\Pi_{c(f)}^B+\Pi_{c(f)}^F$ as in
  Eq.\ (\ref{pi00}), where $\Pi^F_{c(f)}$ is regular in the limit $k\rightarrow k_F$ and $\omega\rightarrow0$ and
    \begin{eqnarray}
    \label{picf}
    \Pi_{c(f)}^B(k, i\omega) & = & \int{d^2q\over(2\pi)^2}n_B(\varepsilon_{\vec{q}})G_{c(f)}(\vec{k}-\vec{q},i\omega-\varepsilon_{\vec{q}})
      \\ \nonumber
      & \sim & xG_{c(f)}(\vec{k},i\omega)+O(\beta^{-1})
   \end{eqnarray}
    where we have followed the same analysis for Eq.\ (\ref{piB0}) and \ (\ref{piB1}) to reach the last result.

   Therefore in the limit $\beta^{-1}\rightarrow0$, as far as the leading contributions from $q\rightarrow0$ is concerned, Eq.\ (\ref{sce})
    can be approximated by
 \begin{eqnarray}
 \label{sce1}
 G_c(k) &=& \frac{-z}{U(1+xU_{eff}G_f(k))} \\
 \nonumber
 G_f(k) &=&
 \frac{1}{i\omega-\bar{\xi}_{\vec{k}}^f-xU^{2}G_c(k)}
 \end{eqnarray}
  where $U_{eff}=U/(1+U\Pi_c^F(k_F,0))$ is the approximate $T$-matrix for bosons scattering with fermions on the Fermi surface evaluated in the
  absence of other bosons. $z=U_{eff}/U$ and
  $\bar{\xi}_{\vec{k}}=\varepsilon_{\vec{k}}+U^2\Pi_f^F(k_F,0)-\mu_f$.

\begin{figure}[h]
\includegraphics [width=9cm]{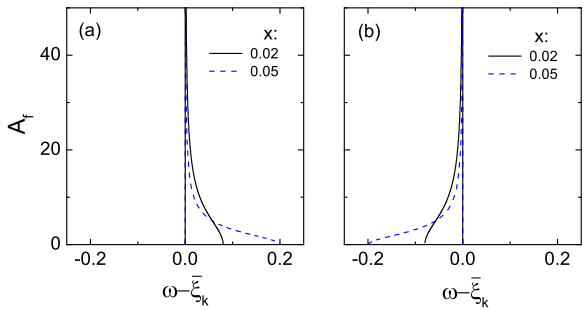}
 \caption{The spectral function $A_f(\vec{k},\omega)$ of the fermions.(a) For $U_{eff}=-1$; (b) for $ U_{eff} = 1 $ for two different values
 of $x=0.02$ and $0.05$. } \label{A}
\end{figure}

 Eq.\ (\ref{sce1}) can be solved easily since it represents quadratic equations for $G_c$ and $G_f$ when the two equations
 are combined. There are two branches of solutions and we pick up the one which reproduces non-interacting fermions in the limit
 $x\rightarrow0$. We obtain,
 \begin{eqnarray}
 \label{gr}
 G_f(\vec{k},i\omega) & = & \frac{1}{\Gamma}\left(\sqrt{1+\frac{2\Gamma}{i\omega-\bar{\xi}_{\vec{k}}}}-1\right)
  \\ \nonumber
 G_c(\vec{k},i\omega) & = &
 -z\left(\frac{i\omega-\bar{\xi}_{\vec{k}}}{U}\right)G_f(\vec{k},i\omega).
\end{eqnarray}
 where $\Gamma=2xU_{eff}$. Notice that $G_f(\vec{k},\omega)\rightarrow(\omega-\bar{\xi}_{\vec{k}})^{-\frac{1}{2}}$ as
 $\omega\rightarrow\bar{\xi}_{\vec{k}}^{f}$, and \emph{there is no pole in the f-fermion Green's function}, indicating that the fermions are in a
 non-Fermi liquid state. The spectral function of the $f-$fermion green's function is
 \begin{eqnarray}
 A_f(\vec{k},\omega)=-\frac{1}{\pi}Im G_{f}(\vec{k},\omega)=\frac{1}{\pi|\Gamma|
}\sqrt{1+\frac{2\Gamma}{\omega-\bar{\xi}_{\vec{k}}}}
\end{eqnarray}
 for $\bar{\xi}_{\vec{k}}-\Gamma-|\Gamma|<\omega<\bar{\xi}_{\vec{k}}-\Gamma+|\Gamma|$
 and $A_f(\vec{k},\omega)=0$ otherwise. The non-Fermi liquid nature of the fermions is obvious as can be seen from Figure \ref{A}.
 It is straightforward to show that the spectral function satisfies the sum rule $\int_{-\infty}^{+\infty} A_f(\vec{k},\omega)d\omega =1$

  Notice that the corresponding composite spectral function $A_c(\vec{k},\omega)=-\frac{1}{\pi}Im G_{c}(\vec{k},\omega)\sim
 |\omega-\bar{\xi}_{\vec{k}}|A_f(\vec{k},\omega)$ is nonzero at the same range of frequencies where $A_f(\vec{k},\omega)$ is
 nonzero. Moreover, $A_c(\vec{k},\omega)$ has no peak and is quite {\em structureless}, indicating that the non-Fermi liquid behavior is
 {\em not} coming from formation of stable, sharp fermion-boson bound states, but is a particular feature of fermion-boson mixture in two
 dimensions.

  The mean-field free energy density of the Bose-Fermi mixture at $T\rightarrow0$ can be computed rather straightforwardly with the approximated Green's
  function\ (\ref{gr}). After some algebra (see supplementary materials)
  we obtain $f(T\rightarrow0)=F(T\rightarrow0)/V=f_M(T\rightarrow0)+f_B(T\rightarrow0)+f_F(T\rightarrow0)$,
  where $f_B$ and $f_F$ are the free energy densities for the corresponding non-interacting Bose and fermi
  gases, respectively,
  \begin{subequations}
  \begin{equation}
  \label{fenergy}
  f_M(T\rightarrow0)=-{\rho\over2}\left({\Gamma^2\over2}+\Gamma\mu_f\right)
  \end{equation}
  is the mean-field free energy coming from bose-fermion interaction, where $\rho=m_f/2\pi$ is the (2D) fermion density of states. Using the
  thermodynamics equality $n_f=-\partial f_M/\partial\mu_f$, we find that the density of fermions is given by
  $n_f=\rho(\mu_f+{\Gamma\over2})$. The corresponding ground state energy density is
  \begin{equation}
  \label{eg}
  \varepsilon_g=\varepsilon_b+\varepsilon_f-U_{eff}n_fx-{\rho\over2}(U_{eff}x)^2,
  \end{equation}
   \end{subequations}
  where $\varepsilon_b$ and $\varepsilon_f$ are the ground state energies for the corresponding non-interacting boson and fermion gases, respectively.
  $U_{eff}n_fx$ is the usual Hartree energy. Our self-consistent theory introduces an additional energy correction $-{\rho\over2}(U_{eff}x)^2$
  corresponding to an effective attractive interaction $\sim-\rho U_{eff}^2$ between bosons mediated through fermions.

\begin{figure}[h]
 \includegraphics [width=9cm]{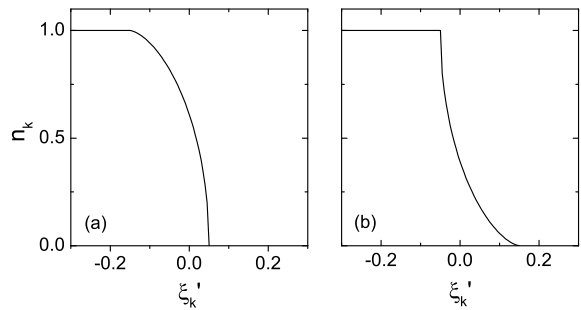}
 \caption{The occupation number of the fermions in momentum space $n_{\vec{k}}$: (a) For $U_{eff} = -1$; (b)for $U_{eff} = 1$. x=0.05. The same density of
 fermions are assumed in both cases.} \label{nk}
 \end{figure}

  The $T\rightarrow0$ fermion occupation number in momentum space $n_{\vec{k}}$ can be calculated by $n_{\vec{k}}=\partial f/\partial\xi_{\vec{k}}$.
  $n_{\vec{k}}$ is shown in figure\ (\ref{nk}) for two different values of $U_{eff}x=\pm0.05$ with fix fermion density $n_f$.
  There is no discontinuity in $n_{\vec{k}}$ across $\bar{\xi}_{\vec{k}}=0$ and the non-Fermi liquid nature of the system is obvious.
  Notice that for fixed fermion number $n_f$, $\mu_f$
  is reduced (increased) from its non-interacting value when $U_{eff}$ is attractive (repulsive).

 %Let $k_F$ be the Fermi momentum for the corresponding non-interacting fermions. We must have $k_{F1}<k_F<k_{F2}$ if the interacting fermion-boson system
 %carries the same total number of fermions as the non-interacting fermions. Therefore the fermi volume shrinks for attractive interaction and expands for
 %repulsive interaction, indicating again the non-perturbative nature of this non-Fermi liquid state.

\subsubsection{interacting bosons}
  In the presence of boson-boson interaction the bosons cannot be integrated out exactly. We shall make an approximation that after integrating out the
  bosons the effective action can still be approximated by $S_{eff}$ (Eq.\ (\ref{seff})), except that the non-interacting boson Green's function $g_b(q)$ is
  replaced by the corresponding Boson Green's function $G_b(q)$ in the presence of interaction, i.e. we still assume a self-consistent Ladder diagram
  approximation in evaluating the effect of fermion-boson interaction.

   Assuming that the interaction between bosons are weak and $S_b$ can be approximated by the usual phase action $S_{\theta}$,
   where the boson operators are approximated by
   \[
     b(\vec{x})\sim\sqrt{\rho_s}e^{i\theta(\vec{x})}, \]
  where $\rho_s$ is the superfluid density, we obtain at low temperature $T<<T_{KT}$, where $T_{KT}$ is the Kosterlitz-Thouless
  transition temperature\cite{popov},
  \begin{equation}
  \label{bboson}
  G_b(\vec{q},\Omega)\sim\rho_s{\delta(\Omega)\over|\vec{q}|^{2-\alpha}}+G_{reg}(\vec{q},\Omega),
  \end{equation}
  where $\alpha=m_b/(2\pi\rho_s\beta)$ and $G_{reg}(\vec{q},\Omega)$ is a regular function in the limit
  $\vec{q},\Omega\rightarrow0$.

  Putting Eq.\ (\ref{bboson}) into Eq.\ (\ref{pis}), we obtain
  $\Pi_{c(f)}(k)=\Pi_{c(f)}^S(k)+\Pi_{c(f)}^{R}(k)$, where
   \[
    \Pi_{c(f)}^{R}(k)={-1\over\beta N}\sum_qG_{f(c)}(k-(+)q)G_{reg}(q)  \]
   is regular on the Fermi surface and
    \begin{equation}
    \label{pii}
    \Pi_{c(f)}(\vec{k},\omega)=\rho_s\int{d^2q\over(2\pi)^2}{1\over|\vec{q}|^{2-\alpha}}G_{f(c)}(\vec{k}-(+)\vec{q},\omega).
    \end{equation}
    We note that $\alpha\rightarrow0$ in the limit $T\rightarrow0$, and the integral has the same logarithmic
    divergence at $\vec{q}\rightarrow0$ as in Eq.\ (\ref{piB0}) or\ (\ref{picf}), indicating that we can extract the leading divergence at
    $\vec{q}\rightarrow0$ in the same way as before, i.e.
    \[
      \Pi_{c(f)}(\vec{k},\omega)\rightarrow\rho_sG_{f(c)}(\vec{k},\omega)+O(\beta^{-1}).
      \]
   suggesting that the same non-Fermi liquid behavior will be obtained as before except $x\rightarrow\rho_s$ in the presence of boson-boson
   interaction. The ground state energy is of the same form as in equation\ (\ref{eg}) except $x\rightarrow\rho_s$ and
   $\varepsilon_b\rightarrow \varepsilon_b^I=$ ground state energy of the corresponding interacting bosons.

 \subsubsection{discussions}
    By summing the leading infra-red singularities in a self-consistent ladder diagram approximation, we show in this paper the existence of a
    $T\rightarrow0 (k_BT>>\Delta\varepsilon)$ non-Fermi liquid state in Bose-Fermi mixtures in two dimensions. The non-Fermi liquid state is a consequence of
    the absence of Bose-condensation (for non-interacting bosons) in two dimension at temperatures $k_BT>>\Delta\varepsilon$ and may persist even when bosons
    are interacting. The true Fermi liquid ground state is recovered only at $k_BT\leq\Delta\varepsilon$, where we may set $n_B(\varepsilon)\sim0$. The
    non-Fermi liquid state we obtained is characterized by a fermion spectral function with no poles, with absence of sharp Fermi surface
    in the corresponding fermion occupation number $n_{\vec{k}}$. The change in shape of the fermion occupation number
    $n_{\vec{k}}$ as the sign of fermion-boson interaction changes\cite{best} may be used as a experimental indicator for this non-Fermi liquid state
    in cold atom systems.
    The non-Fermi liquid state we obtained has lower energy then usual Hartree/Bogoliubov mean-field states and gives rise to effective attractive
    interaction $\sim\rho U_{eff}^2$ between bosons. This is in qualitative agreement with the result obtained in Ref.\cite{zhang} that a
    sufficiently strong boson-boson repulsion is needed to stabilize the Bose-Fermi mixture for both repulsive and attractive Bose-Fermi interaction.

    It should be cautioned that our result is obtained from a particular form of mean-field theory. The range of validity of the mean-field
    theory is unclear and can be addressed only through a careful Renormalization Group (RG) analysis which is out of the scope of the present paper. Nevertheless, our
    analysis demonstrates that the absence of Bose-condensation in two dimension may have a profound effect on Bose-Fermi systems and may lead to
    unconventional states not covered by conventional mean-field theories. A more rigorous theoretical analysis of the problem
    will be carried out in a future paper.

      {\em Acknowledgement} T.K. Ng acknowledges helpful discussions with T.-L. Ho, S.
      Zhang and Q. Zhou. X.Y. Feng acknowledge support of the work in part by the NSF of
      China under grant 11304071.

\begin{widetext}
\section*{Supplementary Material}
 In this Supplementary Material, we provide the derivation of the free energy and the occupation number of the fermions $n_f(\vec{k})$ for the
 bose-fermi mixtures in our mean-field theory.

\subsection*{A. The free energy}
 We start from the expression for the mean-field free energy
 term,
\begin{eqnarray}
F=-\frac{1}{\beta
V}\sum_{k=\vec{\textbf{k}},i\omega}[\ln{G_f^{-1}(k)}+\ln{G_{c}^{-1}(k)}]+\frac{U_{0}^{2}}{\beta
V}\sum_{k=\vec{\textbf{k}},i\omega}\Pi_{f}(k)G_f(k).
\end{eqnarray}
 Since $\Pi_{f}(k)\sim xG_c(k)$ ($k=(\vec{k},i\omega)$) and $G_c(k)=-\frac{z}{U_0g_0(k)}G_f(k)$ where $g_0(k)=(i\omega-\bar{\xi}_{\vec{\textbf{k}}})^{-1}$
 (Eq.(10) in main text), we have
\begin{eqnarray}
F\sim-\frac{1}{\beta
V}\sum_{k=\vec{\textbf{k}},i\omega}[2\ln{G_f^{-1}(k)}-\ln{g_{0}^{-1}(k)}]-\frac{zU_{0}x}{\beta
V}\sum_{k=\vec{\textbf{k}},i\omega}\frac{G_f(k)^2}{g_0(k)}
\end{eqnarray}
 where we have dropped a $\beta$ independent constant which has no thermodynamical effect.

We first consider $F_f=-\frac{1}{\beta
V}\sum_{k=\vec{\textbf{k}},i\omega}\ln{G_f^{-1}(k)}$. Writing
$\ln{G_f^{-1}(k)}=-\frac{1}{\pi}\int_{-\infty}^{\infty}\frac{A(\vec{\textbf{k}},\varepsilon)d\varepsilon}{i\omega-\varepsilon}$
where
\begin{eqnarray}
A(\vec{\textbf{k}},\varepsilon)=Im(\ln{[G_f^{-1}(\vec{\textbf{k}},\varepsilon+i\delta)]})=-\tan^{-1}{\left(\frac{Im{G_f(\vec{\textbf{k}},\varepsilon+i\delta)}}{Re{G_f(\vec{\textbf{k}},\varepsilon+i\delta)}}\right)}
\end{eqnarray}
We have $F_f =
-\frac{1}{V}\sum_{\vec{\textbf{k}}}\int_{-\infty}^{\infty}\frac{d\varepsilon}{\pi}n_f(\varepsilon)\tan^{-1}{\left(\frac{Im{G_f(\vec{\textbf{k}},\varepsilon+i\delta)}}{Re{G_f(\vec{\textbf{k}},\varepsilon+i\delta)}}\right)}$
where $n_f(\varepsilon)=(e^{\beta\varepsilon}+1)^{-1}$ is the Fermi
distribution function.

 Using $ G_f(k) = \frac{1}{\Gamma}(\sqrt{1+2\Gamma g_0(k)}-1)$, we
 obtain also
\begin{eqnarray}
\frac{z_cU_{0}x}{\beta
V}\sum_{k=\vec{\textbf{k}},i\omega}\frac{G_f(k)^2}{g_0(k)}\sim-\frac{1}{\beta
V}\sum_{k=\vec{\textbf{k}},i\omega}\frac{G_f(k)}{g_0(k)}=\frac{1}{V}\sum_{\vec{\textbf{k}}}\int_{-\infty}^{\infty}\frac{d\varepsilon}{\pi}n_f(\varepsilon)Im\left(\frac{G_f(\vec{\textbf{k}},\varepsilon+i\delta)}{g_0(\vec{\textbf{k}},\varepsilon+i\delta)}\right)
\end{eqnarray}

Therefore
\begin{eqnarray}
F =
-\frac{2}{V}\sum_{\vec{\textbf{k}}}\int_{-\infty}^{\infty}\frac{d\varepsilon}{\pi}n_f(\varepsilon)\tan^{-1}{\left(\frac{Im{G_f(\vec{\textbf{k}},\varepsilon+i\delta)}}{Re{G_f(\vec{\textbf{k}},\varepsilon+i\delta)}}\right)}-\frac{1}{V}\sum_{\vec{\textbf{k}}}\int_{-\infty}^{\infty}\frac{d\varepsilon}{\pi}n_f(\varepsilon)Im\left(\frac{G_f(\vec{\textbf{k}},\varepsilon+i\delta)}{g_0(\vec{\textbf{k}},\varepsilon+i\delta)}\right)-F_{0f}
\end{eqnarray}
 where $F_{0f} = \frac{1}{\beta V}\sum_{k=\vec{\textbf{k}},i\omega}\ln{g_0^{-1}(k)}$
($=-\frac{m_f}{2\pi}\frac{\mu^2}{2}$ at zero temperature) is the
free energy for free fermions.

 Notice that $\bar{\xi}_{\vec{\textbf{k}}} = \frac{\vec{\textbf{k}}^{2}}{2m_f} -\mu$ where
 $\mu=\mu_f-U_0^2\Pi_f^F(k_F,0)$ and $d\bar{\xi}=kdk/m_f$, we may change the summation over $\vec{k}$ from $d^2k$ to $d\bar{\xi}$,
 and the free energy of the system becomes,
\begin{eqnarray}
F =
-\frac{m_f}{2\pi}\int_{-\infty}^{\infty}\frac{d\varepsilon}{\pi}n_f(\varepsilon)
\int_{-\mu}^{\infty}d\bar{\xi}\left[2\tan^{-1}{\left(\frac{Im{G_f(\vec{\textbf{k}},\varepsilon+i\delta)}}{Re{G_f(\vec{\textbf{k}},
\varepsilon+i\delta)}}\right)}+Im\left(\frac{G_f(\vec{\textbf{k}},\varepsilon+i\delta)}{g_0(\vec{\textbf{k}},\varepsilon+i\delta)}\right)\right]-F_{0f}.
\end{eqnarray}

 Denoting $x=\varepsilon-\bar{\xi}+\Gamma$, It is straightforward to show that
 $Im\left(\frac{G_f(\vec{\textbf{k}},\varepsilon+i\delta)}{g_0(\vec{\textbf{k}},\varepsilon+i\delta)}\right)=-\sqrt{1-\frac{x^2}{\Gamma^2}}$
 when $|x|<|\Gamma|$ and is equal to zero otherwise.  We have also
\begin{eqnarray}
\begin{array}{ccccccccc}
  ImG_f(\vec{\textbf{k}},\varepsilon+i\delta)&=& -\frac{1}{|\Gamma|}\sqrt{\frac{\Gamma+x}{\Gamma-x}}, & ReG_f(\vec{\textbf{k}},\varepsilon+i\delta)&=&-\frac{1}{\Gamma}, & (|x|&\leq&|\Gamma|) \\
  ImG_f(\vec{\textbf{k}},\varepsilon+i\delta)&=& -\frac{\delta}{|x-\Gamma|\sqrt{x^2-\Gamma^2}}, & ReG_f(\vec{\textbf{k}},\varepsilon+i\delta) &=& \frac{1}{\Gamma}\left(\sqrt{\frac{x+\Gamma}{x-\Gamma}}-1\right), &
  (|x|&>&|\Gamma|)
\end{array},
\end{eqnarray}
 and
\begin{eqnarray}
\tan^{-1}{\left(\frac{Im{G_f(\vec{\textbf{k}},\varepsilon+i\delta)}}{Re{G_f(\vec{\textbf{k}},\varepsilon+i\delta)}}\right)}=\left\{\begin{array}{ccccc}
                                                                                                                            \tan^{-1}{\left(\frac{\Gamma}{|\Gamma|}\sqrt{\frac{\Gamma+x}{\Gamma-x}}\right)}, & (|x|&\leq&|\Gamma|) \\
                                                                                                                              \pi, & (x&>&|\Gamma|)\\
                                                                                                                              0,&
                                                                                                                              (x&<&-|\Gamma|)
                                                                                                                               \end{array}\right..
\end{eqnarray}

 Changing the integration variable from $\bar{\xi}$ to $x$ and considering the case of zero
 temperature, we obtain
\begin{eqnarray}
F &=&
-\frac{m_f}{2\pi}\int_{-\infty}^{0}\frac{d\varepsilon}{\pi}\int_{-\infty}^{\varepsilon+\Gamma+\mu}dx\left[2\tan^{-1}{\left(\frac{Im{G_f(\vec{\textbf{k}},\varepsilon+i\delta)}}{Re{G_f(\vec{\textbf{k}},\varepsilon+i\delta)}}\right)}+Im\left(\frac{G_f(\vec{\textbf{k}},\varepsilon+i\delta)}{g_0(\vec{\textbf{k}},\varepsilon+i\delta)}\right)\right]-F_{0f}\nonumber\\
&=&-\frac{m_f}{2\pi}\int_{-\infty}^{0}\frac{d\varepsilon}{\pi}\left\{\theta(\varepsilon+\mu+\Gamma+|\Gamma|)\int_{-|\Gamma|}^{\min{(\varepsilon+\Gamma+\mu,|\Gamma|)}}dx\left[2\tan^{-1}{\left(\frac{\Gamma}{|\Gamma|}\sqrt{\frac{\Gamma+x}{\Gamma-x}}\right)}-\sqrt{1-\frac{x^2}{\Gamma^2}}\right]\right.\nonumber\\
&&\left.+\theta(\varepsilon+\mu+\Gamma-|\Gamma|)\int_{|\Gamma|}^{\varepsilon+\mu+\Gamma}dx2\pi\right\}-F_{0f}.
\end{eqnarray}

 It is straightforward to obtain after performing the integrals,
\begin{eqnarray}
F=F_{0f}-\frac{m_f}{2\pi}\left(\frac{\Gamma^2}{4}+\frac{\Gamma\mu}{2}\right)=-\frac{\rho}{2}\left(\frac{\Gamma^2}{2}+\Gamma\mu+\mu^2\right)
\end{eqnarray}
%In the limit $U_0x\rightarrow0$,
%\begin{eqnarray}
%G_f(\vec{\textbf{k}},\omega)=\frac{1}{\Gamma}\sqrt{1+\frac{2\Gamma}{\omega+i\delta-\xi_{\vec{\textbf{k}}}}-1}\approx
%g_0(k)\left(1-\frac{\Gamma}{2}g_0(k)\right)
%\end{eqnarray}
%In this case
%\begin{eqnarray}
%F\sim-\frac{1}{\beta
%V}\sum_{k=\vec{\textbf{k}},i\omega}\left[\ln{g_f^{-1}(k)-2\ln{\left(1-\frac{\Gamma}{2}g_0(k)\right)}}\right]
%+
%\end{eqnarray}

\subsection*{B. The Fermion occupation number}
Using Equation(6) and considering the case of zero temperature, we
have
\begin{eqnarray}
F =
-\frac{m_f}{2\pi}\int_{-\infty}^{0}\frac{d\varepsilon}{\pi}\int_{-\mu}^{\infty}d\bar{\xi}\left[2\tan^{-1}{\left(\frac{Im{G_f(\vec{\textbf{k}},\varepsilon+i\delta)}}{Re{G_f(\vec{\textbf{k}},\varepsilon+i\delta)}}\right)}+Im\left(\frac{G_f(\vec{\textbf{k}},\varepsilon+i\delta)}{g_0(\vec{\textbf{k}},\varepsilon+i\delta)}\right)\right]-F_{0f}
\end{eqnarray}

In the case of $\Gamma>0$, carrying out the integral in
$\varepsilon$ first, the free energy becomes
\begin{eqnarray}
F &=&
-\frac{m_f}{2\pi}\int_{-\mu}^{\infty}d\bar{\xi}\int_{-\infty}^{-\bar{\xi}+\Gamma}\frac{dx}{\pi}\left[2\tan^{-1}{\left(\frac{Im{G_f(\vec{\textbf{k}},\varepsilon+i\delta)}}{Re{G_f(\vec{\textbf{k}},\varepsilon+i\delta)}}\right)}+Im\left(\frac{G_f(\vec{\textbf{k}},\varepsilon+i\delta)}{g_0(\vec{\textbf{k}},\varepsilon+i\delta)}\right)\right]-F_{0f}\nonumber\\
&=&-\frac{m_f}{2\pi}\int_{-\mu}^{0}d\bar{\xi}\int_{-\Gamma}^{\Gamma}\frac{dx}{\pi}\left[2\tan^{-1}{\left(\sqrt{\frac{\Gamma+x}{\Gamma-x}}\right)}-\sqrt{1-\frac{x^2}{\Gamma^2}}\right]\nonumber\\
&&-\frac{m_f}{2\pi}\int_{-\mu}^{0}d\bar{\xi}\int_{\Gamma}^{\Gamma-\bar{\xi}}\frac{dx}{\pi}2\pi-F_{0f}\nonumber\\
&&-\frac{m_f}{2\pi}\int_{0}^{2\Gamma}d\bar{\xi}\int_{-\Gamma}^{\Gamma-\bar{\xi}}\frac{dx}{\pi}\left[2\tan^{-1}{\left(\sqrt{\frac{\Gamma+x}{\Gamma-x}}\right)}-\sqrt{1-\frac{x^2}{\Gamma^2}}\right]\nonumber\\
&=&\frac{m_f}{2\pi}\int_{-\mu}^{0}d\bar{\xi}\left(\bar{\xi}-\frac{\Gamma}{2}\right)+\frac{m_f}{2\pi}\int_{0}^{2\Gamma}d\bar{\xi}\frac{\Gamma}{2\pi}\left[\left(\sin^{-1}{\alpha}+\frac{\pi}{2}\right)(1-2\alpha)+(\alpha-2)\sqrt{1-\alpha^2}\right]\nonumber\\
&=&\frac{1}{V}\sum_{\vec{\textbf{k}}(-\mu<\bar{\xi}_{\vec{\textbf{k}}}<0)}\left(\bar{\xi}_{\vec{\textbf{k}}}-\frac{\Gamma}{2}\right)+\frac{1}{V}\sum_{\vec{\textbf{k}}(0<\bar{\xi}_{\vec{\textbf{k}}}<2\Gamma)}\frac{\Gamma}{2\pi}\left[\left(\sin^{-1}{\alpha}+\frac{\pi}{2}\right)(1-2\alpha)+(\alpha-2)\sqrt{1-\alpha^2}\right]
\end{eqnarray}
in which
$\alpha=\frac{\Gamma-\bar{\xi}_{\vec{\textbf{k}}}}{\Gamma}$.

 The Fermion occupation number can be determined by the thermodynamics equality $n_{\vec{\textbf{k}}}=V\frac{\partial
F}{\partial\bar{\xi}_{\vec{\textbf{k}}}}$. We obtain,
\begin{eqnarray}
n_{\vec{\textbf{k}}}=\left\{\begin{array}{cc}
                       1, & \bar{\xi}_{\vec{\textbf{k}}}<0 \\
                       \frac{1}{\pi}\left(\sin^{-1}{\alpha}+\frac{\pi}{2}-\sqrt{1-\alpha^2}\right), &
                       0<\bar{\xi}_{\vec{\textbf{k}}}<2\Gamma\\
                       0,&\bar{\xi}_{\vec{\textbf{k}}}\geq2\Gamma
                     \end{array}\right.
\end{eqnarray}

 For $\Gamma<0$, it is also easy to show that
\begin{eqnarray}
G_f^R(x';-\Gamma)=-G_f^{A}(-x';\Gamma)
\end{eqnarray}
in which $x'$ denotes $\omega - \bar{\xi}_{\vec{\textbf{k}}}$ and
therefore
\begin{eqnarray}
n(\bar{\xi}_{\vec{\textbf{k}}};\Gamma)&=&1-n(-\bar{\xi}_{\vec{\textbf{k}}};-\Gamma)
\end{eqnarray}
We therefore obtain the expression for $\Gamma<0$,
\begin{eqnarray}
n_{\vec{\textbf{k}}}=\left\{\begin{array}{cc}
                       0, & \bar{\xi}_{\vec{\textbf{k}}}>0 \\
                       \frac{1}{\pi}\left(-\sin^{-1}{\alpha}+\frac{\pi}{2}+\sqrt{1-\alpha^2}\right), &2\Gamma<\bar{\xi}_{\vec{\textbf{k}}}<0
                       \\
                       1,&\bar{\xi}_{\vec{\textbf{k}}}<2\Gamma
                     \end{array}\right.
\end{eqnarray}

\end{widetext}
\end{document}